%% file: AB_C_main.tex
\documentclass[10pt]{article}
\usepackage{latexsym}
\usepackage[dvips]{graphicx}
\usepackage{overcite}
\usepackage{amssymb}
\usepackage{amsbsy}
\usepackage{amsmath}
\usepackage{mathrsfs}
\usepackage{xr}
\usepackage{booktabs} 
\usepackage{multirow} 
\usepackage{xcolor}	  
\usepackage{mathtools}
\usepackage{graphicx}
\usepackage{tabularx}
\usepackage[labelformat=simple, font=small,labelfont=bf]{caption}
\usepackage{placeins}
\usepackage{authblk}
\graphicspath{ {./plot/} }

\DeclareRobustCommand*{\Figure}[3]{
   \begin{figure}[!htb]
   \begin{center}
   \noindent
   \includegraphics[width=#2]{#1}  
   \end{center}
   \caption{#3}
   \addtocontents{lof}{\vspace{\baselineskip}}
   \label{fig:#1}
   \end{figure}
}

\newcommand{\be}{\begin{equation}}
\newcommand{\ee}{\end{equation}}

\begin{document}

\title{Tuning Diblock Copolymer Morphology by Adding Associative Homopolymers}

\author[1]{Xiangyu Zhang\footnote{E-mail: xzhan357@jh.edu}}
\author[2]{Jing Zong}
\author[3]{Dong Meng}
\affil[1]{Department of Chemical and Biomolecular Engineering, John Hopkins University, Baltimore, MD 21218, United States}
\affil[2] {Dave C. Swalm School of Chemical Engineering, Mississippi State University, MS 39762, United States}
\affil[3]{Biomaterials Division, Department of Molecular Pathobiology, New York University, New York, NY 10010, United States}

\maketitle

\begin{abstract}
The ability to tune the microstructures formed by block copolymers using accessible physical approaches provides control for practical material applications. A common strategy involves the addition of homopolymers, which can induce morphological changes through their preferential partitioning into specific microdomains. More recently, supramolecular interactions — being chemistry-specific and stimuli-responsive — have emerged as powerful tools for enabling switchable morphologies.
To gain microscopic insight into this process, we present a simulation study of diblock copolymers blended with homopolymers that selectively associate with one of the blocks via reversible associations. By varying the mode of association, we examine the structural changes induced by supramolecular complexation and compare them with those arising from Van der Waals (VDW) interactions. Our results reveal that, despite exhibiting similar levels of homopolymer partitioning, the lamellar structures differ significantly between the association-driven system and the VDW driven system. Cluster analysis indicates that only small clusters form at weak association strength, whereas a continuous network emerges under strong association conditions. Dynamic analysis further indicates that both morphology and supramolecular binding kinetics significantly influence the diffusion of homopolymers across microdomains, highlighting the material’s potential responsiveness to external stimuli.

\end{abstract}

\input{Introduction}
\input{model_method}
\input{Results}

\newpage
\bibliographystyle{unsrt}
\bibliography{citation}

\end{document}

%% file: Introduction.tex
\section{Introduction}
The blending is commonly used in the polymer processing, acting as a simple and effective way to fabricate polymeric materials with advanced mechanical and thermal properties over their individual constituent \cite{tucker2002microstructural}. The resulted tremendous parameter space of blending system can offer improved control of the micro-structure period \cite{stoykovich2005directed}. Because of its excellent properties, like tunability and stimuli-responsiveness, the application of it scans a wide range of areas, such as, drug delivery \cite{yang2015micellar}, photovoltaic material \cite{tan2014thermally}, and so forth \cite{hofman2015poly,tan2015transient}. \par
The most extensively studied blending system is the mixture of A-B diblock copolymers and C homopolymers. Typically, it can provide more fascinating and versatile behaviors than A-B/A or A-B/B blends \cite{chen2009self,tang2010thin,zhao1997binary,kwak2015phase,han2012highly}. The way to modify the system structure can be the change of the block copolymer segment volume fraction, but that involves the re-synthesis of the polymer and it is certainly cumbersome and times-consuming. Alternatively, the introduction of C homopolymer can avoid these problems and impart abundant morphology behaviors \cite{kuo2022hydrogen}. The incompatible A-block and B-block are usually chosen for A-B copolymer, so that it has the tendency to self-assemble into ordered micro-structures, such as lamellar, sphere, cylinder \cite{mai2012self,darling2007directing}. There are two scenarios depending on the choose of the C-homopolymer. First, C-homopolymer is miscible with A-block, but it is immiscible with B-block, such as the blend of poly(styrene-b-vinylphenol) (PS-b-PVPh, A-b-B) and poly(methyl methacrylate) (PMMA, C)\cite{zhao1997binary,kosonen2001self}. Second, C-homopolymer is miscible with both A-block and B-block, such as the blend of PS-b-PVPh (A-b-B) and poly(vinyl methylether) (PVME, C) \cite{zhao1997binary}. In both cases, the mixing, no matter with which block, is basically promoted by the formation of hydrogen bonds, and it has been confirmed that the strength of hydrogen bonds plays an important role in deciding the system behavior \cite{kuo2022hydrogen,tsai2016hydrogen}. Therefore, a model which can capture reversible interactions correctly is needed to provide insights about the hydrogen bond effects. \par
A variety of studies have been done to investigate the self-assembly of copolymers, but the description of the associative interaction, such as hydrogen bonds, is still a challenge \cite{muller2020process}. In the past, the common way to treat hydrogen bonds in self-consistent field theory is to use the negative Flory-Huggins $\chi$ parameter, suggesting the attractive interaction resulted by hydrogen bonds \cite{lefevre2010self}. That makes it difficult to compare with the experimental observation quantitatively due to the reduced segregation caused by negative $\chi$ \cite{lefevre2010self}. And it either cannot describe the saturation of the associative bond, which is a primary difference with Van der Waals-type interactions \cite{rubinstein2003polymer}. Then, association model was proposed to incorporate the concept of donor and acceptor. One example is the assumption of complete linear complexation, assuming the association of C homopolymer and A-block will lead to a new diblock copolymer \cite{dehghan2013modeling}. But it does not take into account the association entropy, or in other words, all possible associative patterns, which not only makes contribution to the free energy but also affects other properties, like dynamics. The Flory-Huggins energy based association model used for A/B homopolymer blends study can alleviate the deviation process including association activities \cite{prusty2018thermodynamics}. But it can only deal with randomly distributed stickers, assuming the distribution of them will not affect the thermodynamics properties. \par
Simulations offer a viable approach to comprehensively study association effects by explicitly incorporating both association and particle-level information. In this study, smart Monte Carlo movements and reversible associations are implemented. The manuscript is organized as following. The effect of association on the segments distribution is analyzed, and it is also found that association can change the lateral dimension lamellar size, which is proved to be caused by the association patterns. At last, the association effect on dynamics properties is discussed. \par

%% file: model_method.tex
\section{Model and Method}
\subsection{System Descriptions}
We consider a system with a composition of $n_{AB}=1800$ number of symmetric diblock copolymers (DBC) A-B of length $N_{A}+N_{B}$, and $n_{C}=500$ number of C homopolymers of length $N_{C}$, with $N_{A}=N_{B}=N_{C}=10$. A-block and B-block have strong incompatibility interactions to self-assemble into lamellar structure. Two scenarios are considered that are distinguished by the manner of interactions between the homopolymer C and the diblock copolymer A-B. In the first scenario (system I), C homopolymer has strong incompatibility interaction with B-block, leading to the aggregation of C in A-domain, while in the second scenario (system II), they are of supramolecular nature that allows reversible associative bonds being formed between the C-chains and A-blocks to induce the aggregation of C in A-domain. The illustration plot for two scenarios is shown in figure~\ref{fig:9_figure_1.png}. \par

\Figure{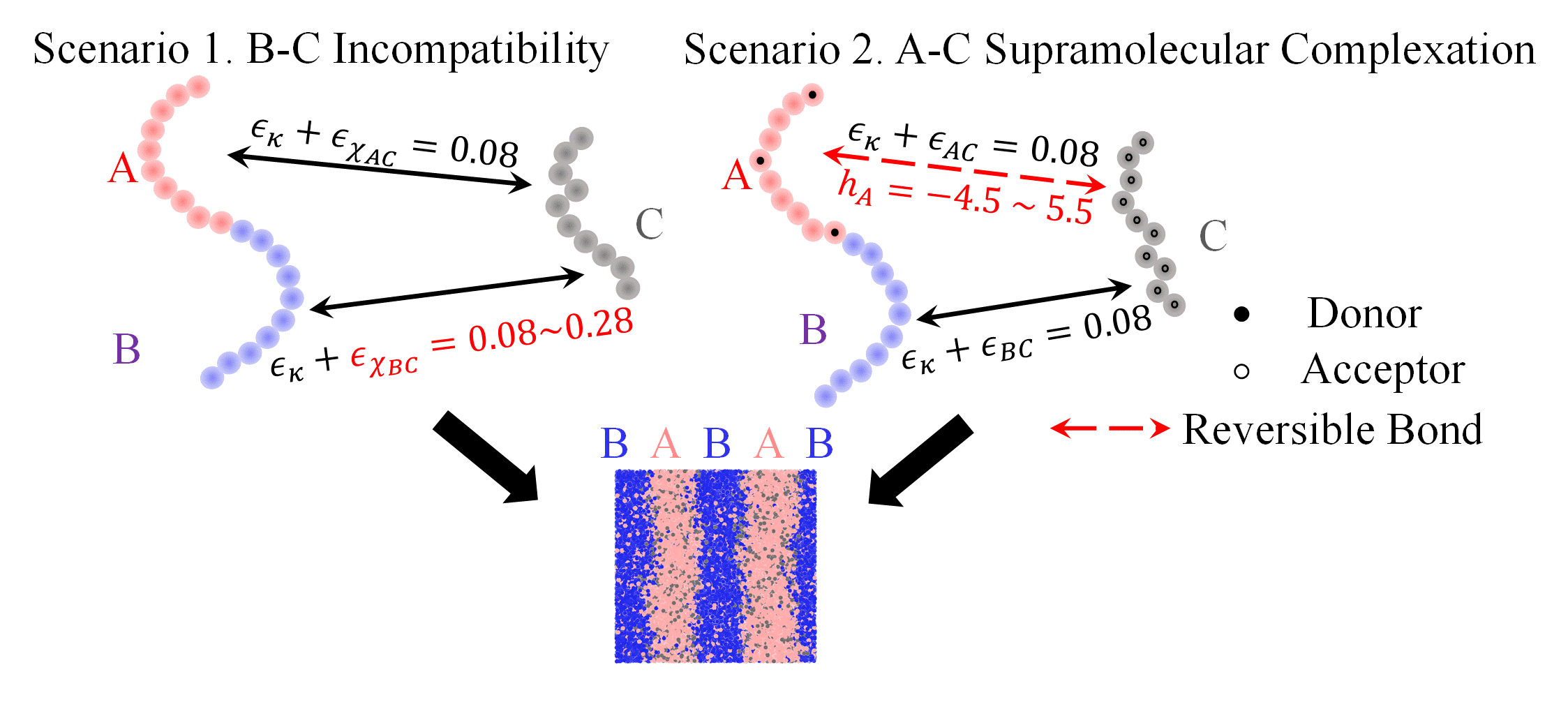}{0.98\linewidth}{Illustration plot of scenario 1, B-block - C-homopolymer incompatibility driven system, and scenario 2, A-block - C-homopolymer supramolecular complexation driven system.}

\subsection{Coarse-Grained Model}
The total free energy of the system can be written as the sum of bonded energy, non-bonded energy and the association energy,
\begin{equation}
\mathcal{H}=\mathcal{H}^{nb}+\mathcal{H}^b+\mathcal{H}^a
\end{equation} 
, where $\mathcal{H}^{nb}$ is the non-bonded energy, $\mathcal{H}^b$ is the bonded energy, and $\mathcal{H}^a$ is the association energy. For covalent systems, the association energy term is equal to $0$. Bonded energy is defined as, 
\begin{equation}
\mathcal{H}^{b}=\sum_{j=1}^{\Phi_{CB}}u^{b}(\left | {\bf r}_{P,j+1} - {\bf r}_{P,j} \right |)
\end{equation}
, where $\Phi_{CB}$ is the total number of covalent bonds in the system, ${\bf r}_{P,j}$ denotes the spatial position of $j$ polymer segment, and $u^{b}$ is the bonding potential, retaining connectivity. In the study, discrete Gaussian bond potential is used,
\begin{equation} \label{eq:DGB_P}
u^{b}(\left | {\bf r}_{P,j+1} - {\bf r}_{P,j} \right |)=\frac{3 k_{B}T}{2a^{2}}\left | {\bf r}_{P,j+1} - {\bf r}_{P,j} \right |^{2}
\end{equation}
, with $a$ being the effective bond length, $k_{B}$ being the Boltzmann constant, and $T$ being the temperature. The total non-bonded energy is given by,
\begin{equation}
\begin{split}
\mathcal{H}^{nb} & = \sum_{i=1}^{n_A\cdot N_A}\sum_{j>i}^{n_A\cdot N_A}\int d{\bf r}\int d{\bf r}' \delta({\bf r}-{\bf r}_{A,i}) u_{AA}^{nb}({\bf r},{\bf r}')\delta({\bf r}'-{\bf r}_{A,j}) \\
& + \sum_{i=1}^{n_B\cdot N_B}\sum_{j>i}^{n_B\cdot N_B}\int d{\bf r}\int d{\bf r}' \delta({\bf r}-{\bf r}_{B,i}) u_{BB}^{nb}({\bf r},{\bf r}')\delta({\bf r}'-{\bf r}_{B,j}) \\
& + \sum_{i=1}^{n_C\cdot N_C}\sum_{j>i}^{n_C\cdot N_C}\int d{\bf r}\int d{\bf r}' \delta({\bf r}-{\bf r}_{C,i}) u_{CC}^{nb}({\bf r},{\bf r}')\delta({\bf r}'-{\bf r}_{B,j}) \\
& +\sum_{i=1}^{n_A\cdot N_A}\sum_{j=1}^{n_B\cdot N_B}\int d{\bf r}\int d{\bf r}' \delta({\bf r}-{\bf r}_{A,i}) u_{AB}^{nb}({\bf r},{\bf r}')\delta({\bf r}'-{\bf r}_{B,j}) \\
& + \sum_{i=1}^{n_A\cdot N_A}\sum_{j=1}^{n_C\cdot N_C}\int d{\bf r}\int d{\bf r}' \delta({\bf r}-{\bf r}_{A,i}) u_{AC}^{nb}({\bf r},{\bf r}')\delta({\bf r}'-{\bf r}_{C,j}) \\
& + \sum_{i=1}^{n_B\cdot N_B}\sum_{j=1}^{n_C\cdot N_C}\int d{\bf r}\int d{\bf r}' \delta({\bf r}-{\bf r}_{B,i}) u_{BC}^{nb}({\bf r},{\bf r}')\delta({\bf r}'-{\bf r}_{C,j})
\end{split}
\end{equation}
, where ${\bf r}_{\alpha,k}$ represents the spatial position of the $\alpha$ type $k$ segment, and
\begin{equation}
u_{\alpha \alpha'}^{nb}({\bf r},{\bf r}') \equiv \epsilon_{\alpha\alpha'} \left( 15/2\pi \right) \left(1 - r/\sigma \right)^2
\end{equation}
for $r<\sigma$, where $\sigma$ is the unit length, or $u_{\alpha \alpha'}^{nb}({\bf r},{\bf r}') = 0$ otherwise. $\epsilon_{\alpha\alpha'}$ controls the interaction strength and has the unit of $k_{B}T$, which is defined as, 
$\epsilon_{\alpha\alpha\prime}\equiv\left\{\def\arraystretch{1.2}\begin{tabular}{@{}l@{\quad}l@{}}
  $\epsilon_{\kappa}$ & if $\alpha=\alpha\prime$ \\
  $\epsilon_{\kappa}+\epsilon_{\chi_{\alpha\alpha'}}$ & if $\alpha \neq \alpha\prime$
\end{tabular}\right.$ . $\epsilon_{\kappa}$ is the excluded volume and $\epsilon_{\chi_{\alpha\alpha'}}$ describes the polymer incompatibility. \par
The total association energy for supramolecular systems can be expressed as,
\begin{equation} \label{eq: asso_total}
\mathcal{H}^{a}=\sum_{j=1}^{N_{asso}}u^{a}(\left | {\bf r}_{Do,j} - {\bf r}_{Ac,j} \right |)) 
\end{equation}
, where $N_{asso}$ is the total number of associated pairs in the system, "Do" corresponds to the donor of the association pair, "Ac" is the acceptor, and $u^{a}(\left | {\bf r}_{Do,j} - {\bf r}_{Ac,j} \right |))$ is,
\begin{equation} \label{eq: asso_pair}
u^{a}(\left | {\bf r}_{Do,j} - {\bf r}_{Ac,j} \right |)) = \frac{3 k_{B}T}{2\sigma^{2}}(\left | {\bf r}_{Do,j} - {\bf r}_{Ac,j} \right |)^{2} + h_{A} 
\end{equation}
, where $h_{A}$ is the association constant to control the association probability. The detail of the implementation of association movements in Monte Carlo simulations can be found in the cited paper \cite{zhang2025investigation}. \par

$\epsilon_{\kappa}$ is set to be a constant in both scenarios, which is $0.08$. In scenario 1, $\epsilon_{\chi_{BC}}$ is varied from $0 \sim 0.2$. In scenario 2, $\epsilon_{\chi_{BC}}$ is set to be zero. There are three donors on A-block chains, and ten acceptors on C homopolymer chains, as it is shown in figure~\ref{fig:9_figure_1.png}. We require that one donor can only associate with one acceptor. The association energy barrier, $h_{A}$, is varied from $-4.5 \sim 5.5$. The simulation is running in NPT ensemble to find the equilibrium lamellar period, and the input pressure is $5.5 k_{B}T/\sigma^{3}$. In the simulation, box size in x and z directions can be varied. And box length in y-direction is fixed as $16 \sigma$. Lamellar structure is along x-direction and there are two periods in the box. The field-accelerated Monte Carlo method is implemented to improve the efficiency by converting the particle-based free energy to field-based free energy. More details can be found in the cited paper \cite{zong2020field}. The smart Monte Carlo movement is implemented to study kinetic properties \cite{allen2017computer}.\par

%% file: results.tex
\subsection{Partitioning of C Segments in A-domain}
The partitioning of C segments in A-domain is examined first in figure~\ref{fig:9_figure_2.png} (a). The bottom axis suggests the association energy barrier, and accordingly, the aggregation of C segments in A-domain can be observed in supramolecular system due to A-C attractive interaction. Similarly, the increase of the B-C repulsive interaction can promote the aggregation of C segments in A-domain. But they do show a slight difference. In B-C incompatibility driven system, the number of C-segments keeps increasing with the strengthening of B-C interaction, because Van der Waals type interaction is non-saturated interaction and the interaction partners are all particles within the interaction range. So, it will only reach the maximum value, that is $5000$, when all of C segments are driven to A-domain. But in supramolecular system, it will reach the plateau at the value around $4500$. Next, let us look at the fraction of associated C segments plot in figure~\ref{fig:9_figure_2.png} (b). It suggests that most of C homopolymer associative sites are occupied. Recalling that we have $n_{AB}=1800$, and there are $3$ donors on A-block chain and $10$ acceptors on C homopolymer chain, indicating there are enough donors on A-block. So, we can say that the extent of aggregation caused by supramolecular interaction is weaker than B-C incompatibility driven system. The above finding suggests the qualitatively different nature of the associative interaction from Van der Waals type force, and in further implying the unsuitable use of negative $\chi$ to model associative interaction \cite{beardsley2019calibration}. \par

\Figure{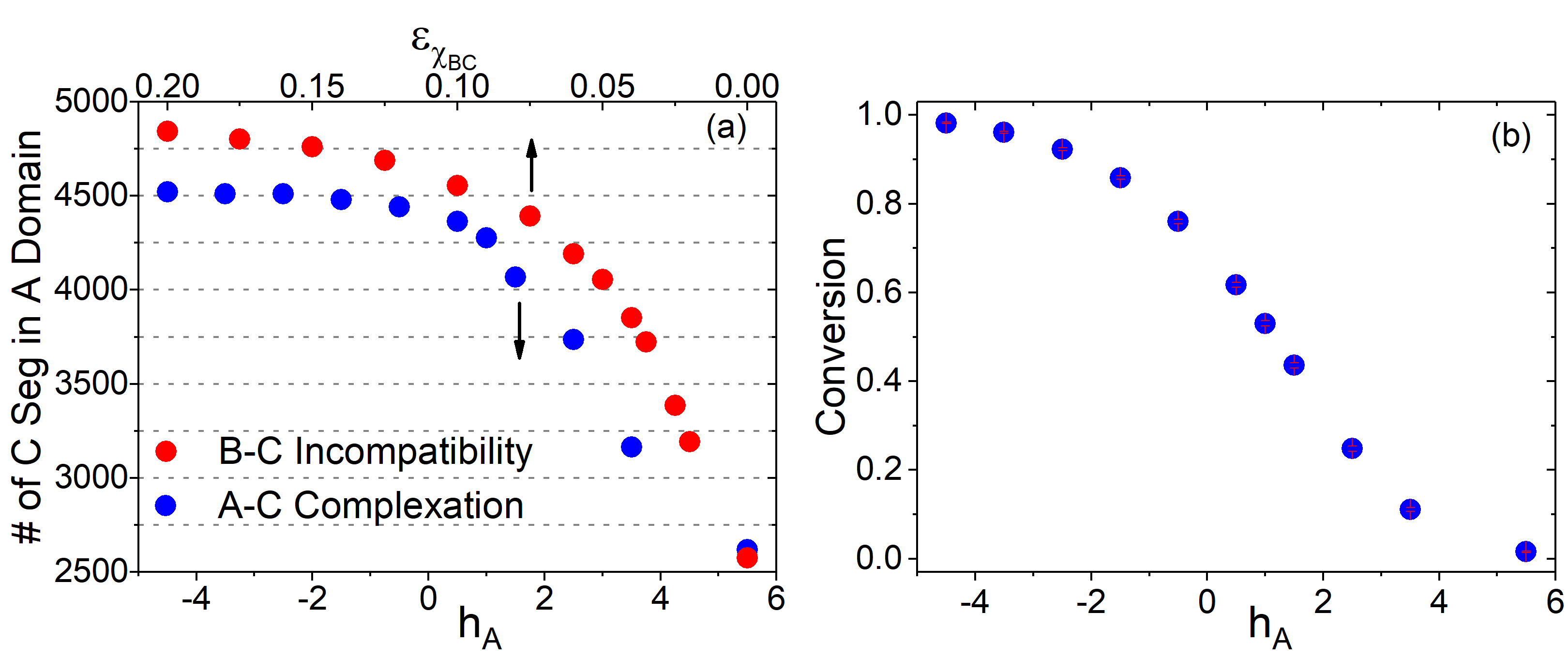}{0.98\linewidth}{(a) The number of C segments in the A-domain as a function of association energy barrier in supramolecular system (bottom axis) and B-C incompatibility system (top axis); (b) The fraction of associated C segments plotted against association energy barrier in supramolecular system.}

The density profile is compared at the same C homopolymer segments partitioning. The bottom axis is the position along lamellar direction, which is x-axis in our simulation, and the left axis is the segments' fraction at that point. In weak C segments segregation system, both scenarios show the similar distribution, as it is shown in figure~\ref{fig:9_figure_3.png}. A peak point can be observed at the junction point between A- and B-block. The reason is the strong incompatibility between A-block and B-block, so, C segments will try to stay at the interface to reduce A-B interfacial energy and in further to minimize the total free energy. Overall, C segments are like the surfactant between two domains. The similar behavior has been reported before for A-B/C blends \cite{shull1992homopolymer}. The significant difference can be observed for strong segregation system in figure~\ref{fig:9_figure_4.png}. The density profile of B-C incompatibility system in figure~\ref{fig:9_figure_4.png} (a) shows the aggregation of C segments in the center of the A-domain. The reason is the repulsive interaction exhibited by C segments from B-blocks. Therefore, C segments will try to stay as far as it can from B-domain, and the furthest place to B-domain will be the center of the A-domain. However, C segments distribute uniformly in the A-domain in association driven system. The reason is that the position of the acceptor on C homopolymer only depends on the position of the donor on A-block. Hence, it does not make any differences for C homopolymers to stay at the junction point or at the center of the domain as long as there is an open donor to associate. The distribution difference of C segments also suggest the importance to incorporate association explicitly and correctly in the model.  \par

\Figure{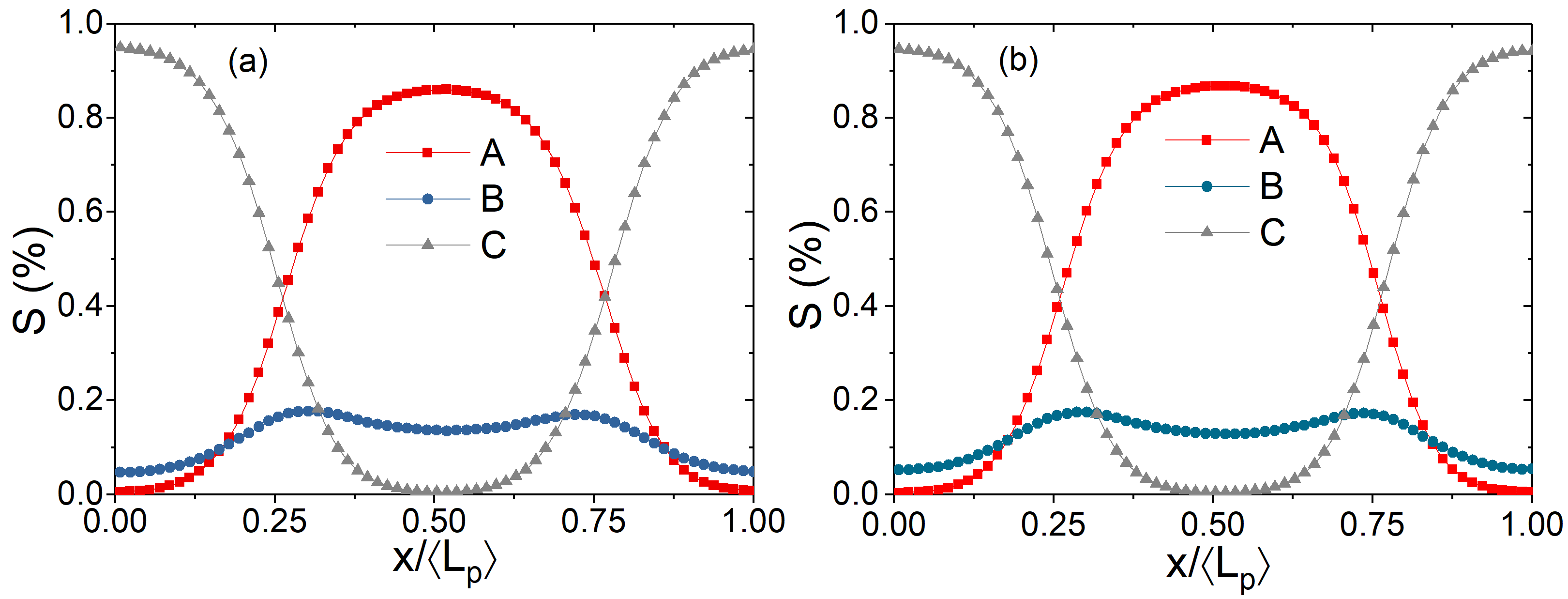}{0.98\linewidth}{The density profile of C segments weak segregation system (number of C segments in A-domain $\approx 3150$) in B-C incompatibility (figure (a), $\epsilon_{\chi_{BC}}=0.02$) driven and  supramolecular complexation driven (figure (b), $h_{A}=3.5$) scenarios.}

\Figure{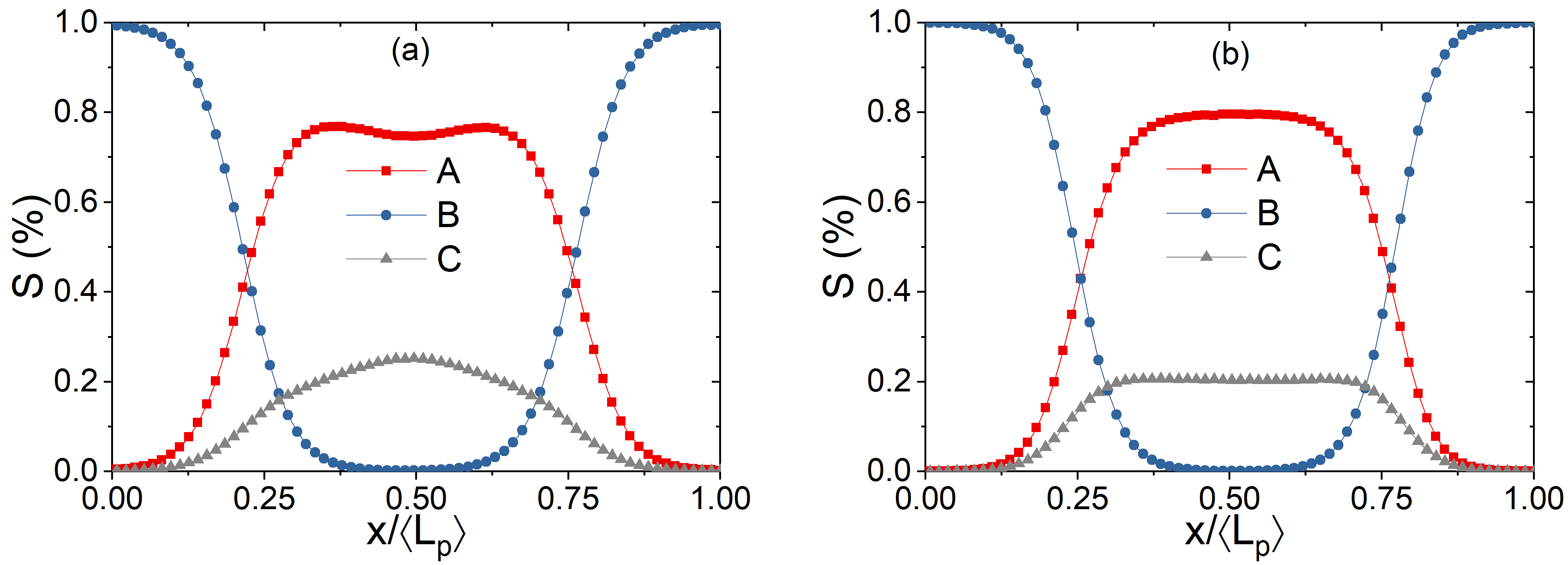}{0.98\linewidth}{The density profile of C segments strong segregation system (number of C segments in A-domain $\approx 4500$) in B-C incompatibility (figure (a), $\epsilon_{\chi_{BC}}=0.1$) driven and  supramolecular complexation driven (figure (b), $h_{A}=-4.5$) scenarios.} 

\clearpage
\subsection{The Association Effect on Lamellar Structure}
The simulation is running in NPT ensemble, so, it allows us to find out the equilibrium box size for the lamellar structure. The lamellar period and the lateral size in figure~\ref{fig:9_figure_5.png} are both normalized by non-preferential (reference) system, that is $\epsilon_{\chi_{BC}}=0$ and $h_{A}=\infty$ system. The equilibrium lamellar and lateral size in reference system are $7.94\sigma$ and $15.62\sigma$, respectively. Because we have two periods, the box length along lamellar direction will be $15.88\sigma$. The expansion of the lamellar period is observed in both supramolecular and incompatibility system in figure~\ref{fig:9_figure_5.png} (a), due to the aggregation of C homopolymers in the A-domain. The degree of increase is similar for both of scenarios, indicating the insignificance of the segments distribution effect on lamellar structure. The change of lateral dimension size, that is the direction perpendicular to the lamellar, is also calculated and it is shown in figure~\ref{fig:9_figure_5.png} (b). 
The system volume is almost a constant upon varying B-C repulsive interaction, which is equal to around $16\times 250 \sigma^{3}$. $16 \sigma$ corresponds to the fixed length in the y-direction. So, the slight decrease of lateral size can be observed for incompatibility system, which can be attributed to the expansion of the lamellar period. In A-C complexation system, the degree of shrinkage is much larger and more pronounced in lateral direction. \par
To figure out what causes this evident decrease, cluster size is analyzed in figure~\ref{fig:9_figure_6.png}. In weak complexation system, that is figure~\ref{fig:9_figure_6.png} (a) plot, the peak is at $n_{AB}=1$, suggesting that there is only one A-B chain in most of association clusters. It indicates that polymer chains are separated from each other. The example of the associative pattern is shown in the inset in the plot. On the other hand, the associative pattern in strong complexation system with average conversion rate being $0.99$ is different. In figure~\ref{fig:9_figure_6.png} (b) plot, there are two groups of points highlighted by red dashed circles. The top axis, representing A–B chain fractions, indicates that these groups are observed simultaneously in a single snapshot, as the overall conversion is 0.99, with respective fractions of approximately 0.46 and 0.53. Additionally, there is one small group of points highlighted by green dashed line, implying that all of A-B chains are included in only one cluster, which can be told by looking at the corresponding fraction on top axis. It can be conjectured that most chains are interconnected and a network has formed, given the presence of only one or two clusters and an overall conversion of $0.99$. Accordingly, the high degree of cross-linking reduces the lateral dimensions of the system, thereby decreasing the associative energy and minimizing the total free energy. This result also suggests the possible mechanical properties change brought by supramolecular interaction due to cross-linking and has been discussed in the cited paper \cite{maitra2014cross,tzoumanekas2006topological}. \par

\Figure{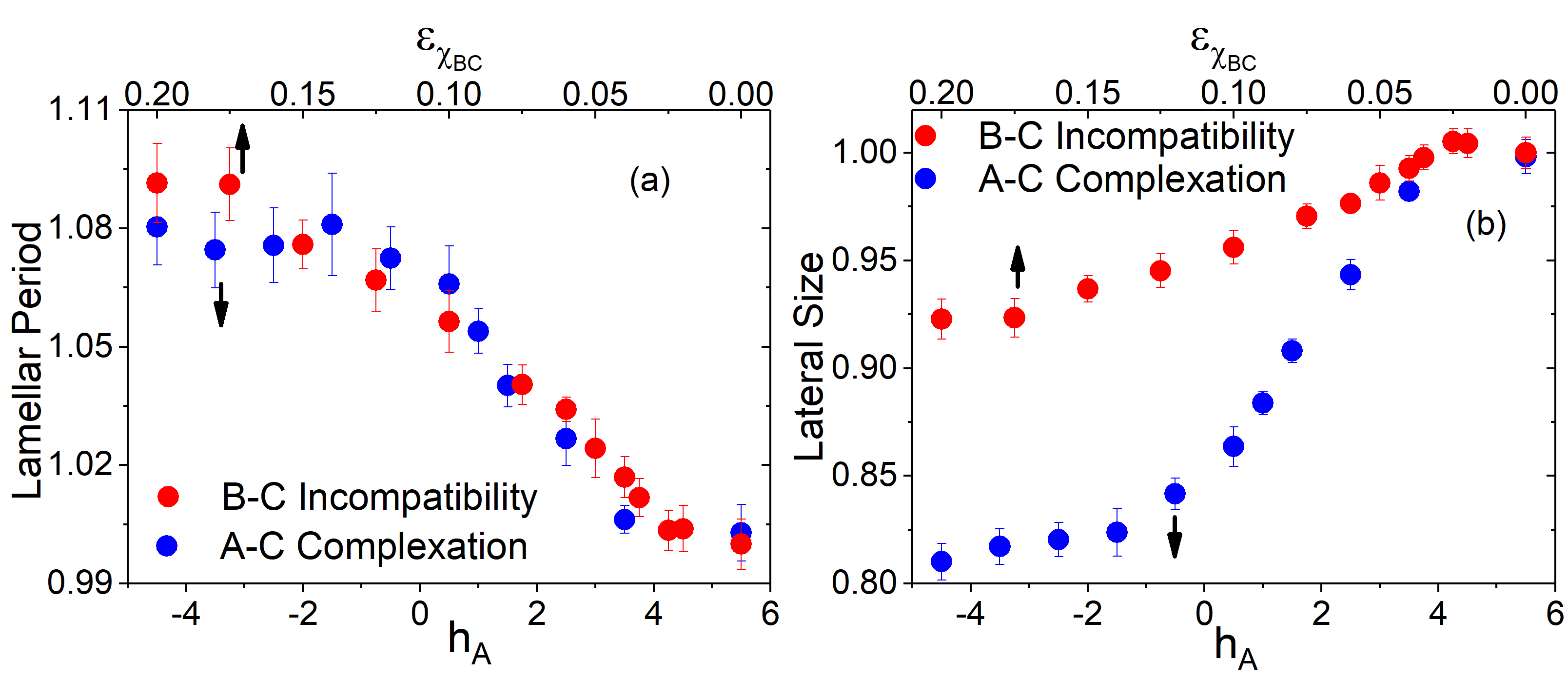}{0.98\linewidth}{(a) Lamellar period normalized by $h_{A}=\infty$ and $\epsilon_{\chi_{BC}}=0$ system as a function of association energy barrier ($h_A$) and B-C repulsive interaction ($\epsilon_{\chi_{BC}}$); (b) Lateral dimension size normalized by $h_{A}=\infty$ and $\epsilon_{\chi_{BC}}=0$ system as a function of association energy barrier and B-C repulsive interaction.}

\Figure{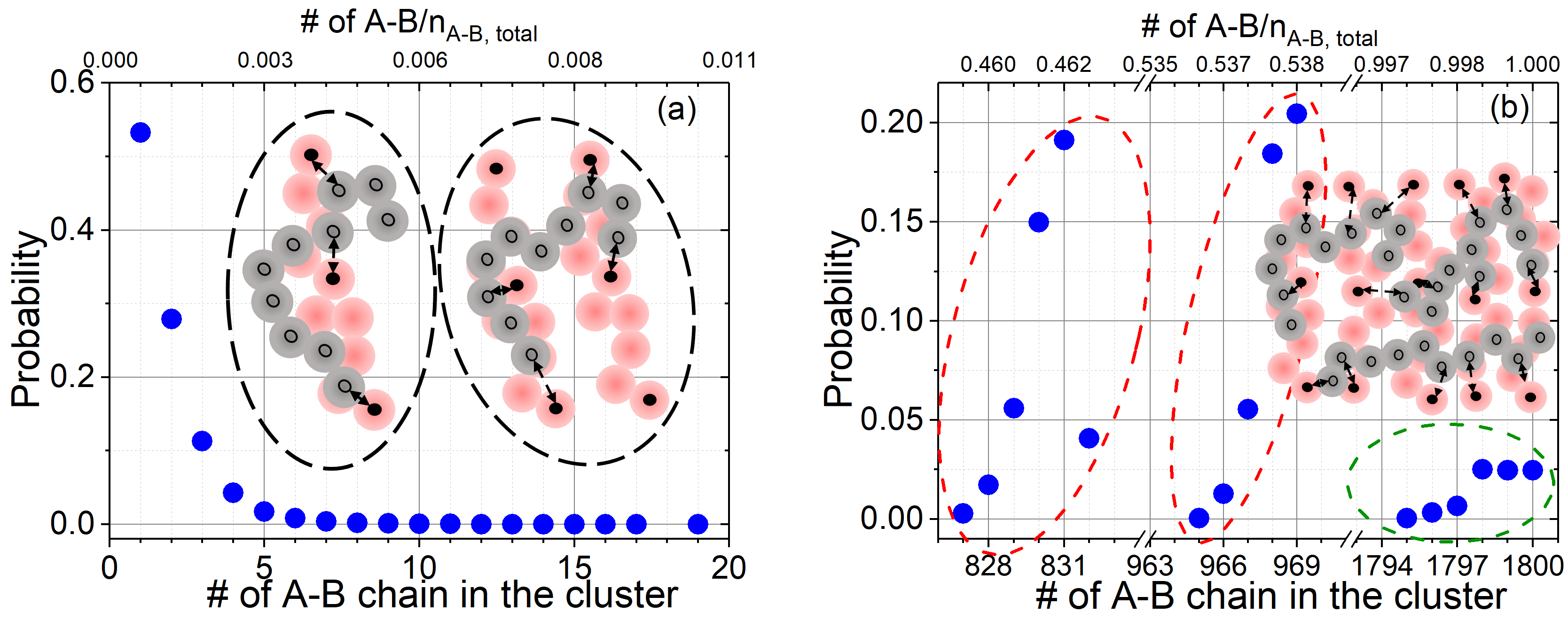}{0.98\linewidth}{(a) The distribution of the number of A-B chains in each cluster with fraction of associated C segments being $0.11$ ($h_A=3.5$). (b) The distribution of the number of A-B chains in each cluster with fraction of associated C segments being $0.99$ ($h_A=-4.5$). For both plots, the top axis shows the corresponding A-B chains fraction.}

\subsection{The Association Effect on Homopolymer Chain Diffusion}

\Figure{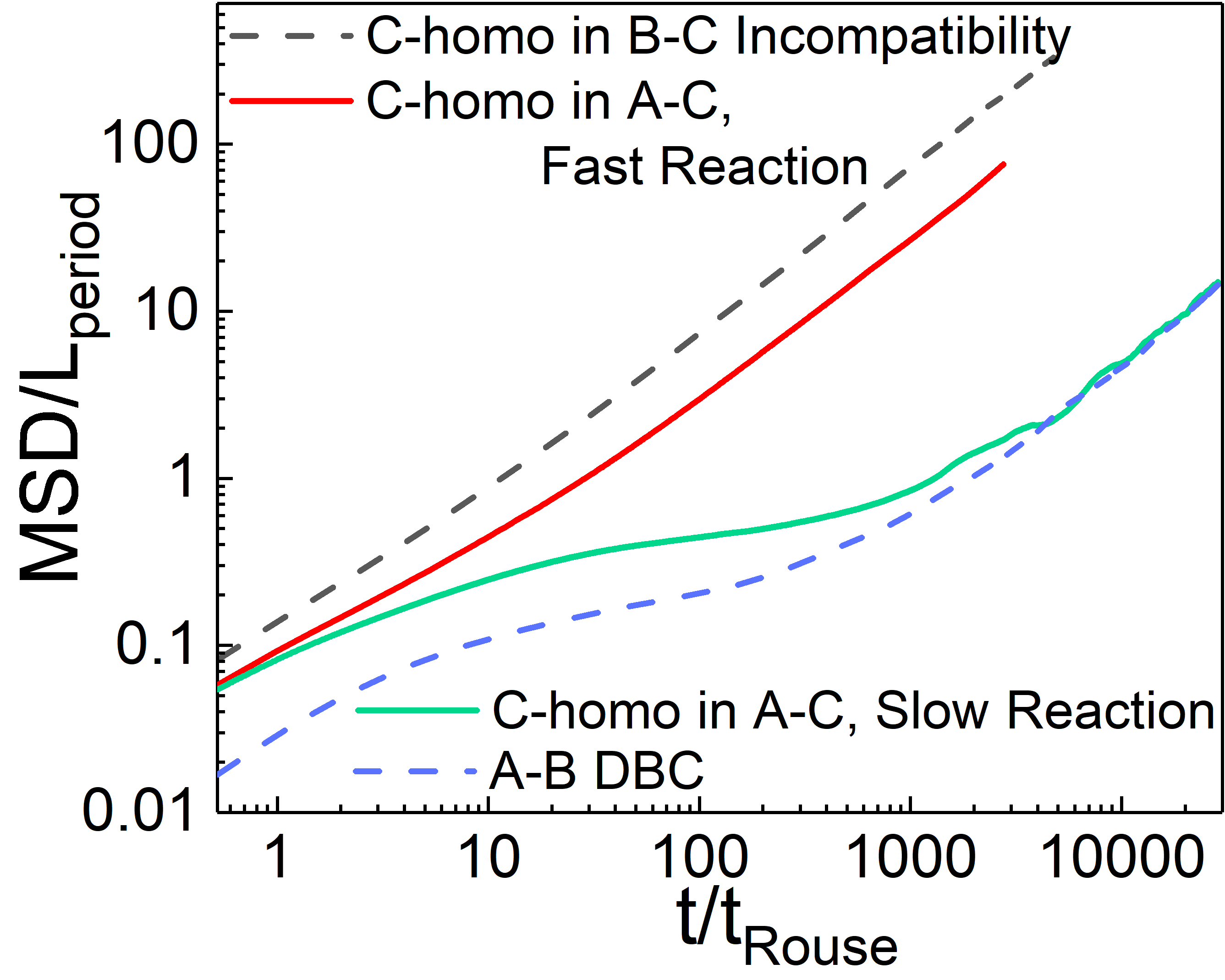}{0.7\linewidth}{The mean-square-displacement of the center of the chain mass normalized by lamellar period plotted against Monte Carlo steps normalized by Rouse time, which is identified to be 2820 MC steps. The blue dashed line is MSD calculated for A-B diblock copolymer, green line corresponding to the diffusion of C homopolymer with reaction rate being extremely slow (bonding forming/breaking probability: $0.00005$), red line representing fast reaction rate system (bonding forming/breaking probability: $0.5$), and grey dashed line suggest the C homopolymer diffusion in non-association system.}

The implementation of smart MC moves allows the capture of Rouse-scale dynamics properties \cite{muller2008single}. The mean-square-displacement (MSD) is plotted in the figure~\ref{fig:9_figure_7.png}. A-B diblock copolymers self-assemble into lamellar structure, so, the diffusion rate of A-B DBC will be the diffusion rate of the system. The extremely slow reaction rate indicates that the lifetime of the associative bonds will be exceptionally long. Once a associative bond is formed between A-block and C homopolymer, C homopolymer will diffuse together with the A-block due to the long-lasting associative bond. Therefore, the blue dashed line, which is the MSD of A-B DBC, becomes the lower limit of the C homopolymer diffusion rate. Clearly, three stages of diffusion can be observed for both green, corresponding to C-homopolymer with slow reaction rate, and blue dashed line, that is A-B DBC. At the start of the curve, it shows a constant slope for a very short of period, that is ballistic regime. Next, the slope of MSD curve begins decreasing until reaching a constant value, corresponding to trapped regime. The chain diffusion is hindered due to the restraint of associative bonds. At last, it will enter the free diffusion regime. The above three regimes are not only for polymer system, but commonly observed in various systems \cite{mahanta2020connection}. If the reaction is very fast, suggesting the short associative bond life. Hence, C homopolymers cannot "feel" the formation of associative bonds, as it will be broken very quickly after the formation. Accordingly, the MSD curve, that is red line, will approach the non-association system, that is B-C incompatibility driven system, corresponding to grey dashed line. So, that is the upper limit of the C homopolymer diffusion rate in A-C supramolecular system. It can also be observed that the trapped regime is hardly appreciable due to short bond life, and the slope in ballistic regime is very close to it in free diffusive regime. \par 

\section{Conclusion}
This work distinguishes the supramolecular effect on the system micro-structure after the equilibrium. The supramolecular complexation can change the distribution of C homopolymer segments, and also the complexation will modify the micro-pattern size. At low degree of complexation system, the behavior is similar for both supramolecular and B-C incompatibility system. But the higher degree of complexation leads to the formation of the network, which in further affects the lateral size of the lamellar. The difference between our full association model and complete linear complexation indicates the importance to incorporate the associative entropy, as it has a direct influence on the micro-structure period. The dynamics study shows the influence of reaction rate, which effectively decides the associative bond life, on the polymer diffusion. It can be simply concluded that the longer the associative bond life, the slower the diffusion rate will be. Both upper and lower C homopolymer diffusion rate limits are characterized by MSD analysis. The upper limit is found to be the non-associative system, corresponding to extremely fast reaction rate system. The lower limit is identified to be the A-B diblock copolymer diffusion rate, as C homopolymer is permanently attached to it due to extremely slow reaction rate. 